# Online platforms of public participation – a deliberative democracy or a delusion?

Jonathan Davies
Department of Computer Science, University of Warwick
Gibbet Hill Road, Coventry, CV4 7AL
Jonathan.Davies.1@warwick.ac.uk

Rob Procter
Department of Computer Science, University of Warwick
& Alan Turing Institute for Data Science
Gibbet Hill Road, Coventry, CV4 7AL
rob.procter@warwick.ac.uk

## ABSTRACT

Trust and confidence in democratic institutions is at an all-time low. At the same time, many of the complex issues faced by city administrators and politicians remain unresolved. To tackle these concerns, many argue that citizens should, through the use of digital platforms, have greater involvement in decision-making processes. This paper describes research into two such platforms, 'Decide Madrid' and 'Better Reykjavik'. Through the use of interviews, questionnaires, ethnographic observation, and analysis of platform data, the study will determine if these platforms provide greater participation or simply replicate what is already offered by numerous other digital tools. The findings so far suggest that to be successful platforms must take on a form of deliberative democracy, allowing for knowledge co-production and the emergence of collective intelligence. Based on this, we aim to identify key features of sustainable models of online participation.

## 1. INTRODUCTION

In recent decades, almost every aspect of life (media, retail, education, tourism, social networks etc.) has been re-shaped, re-formed and re-fashioned by the influence of information and communication technologies (ICTs) [1]. Yet one sphere that seems immune to these transformative effects are democratic institutions, which remain largely unchanged since the 20th century [2, 3]. As a result, a gap between the activities of citizens and the way in which politics and democracy are carried out has developed, contributing to a decline in trust and confidence in governments [4] and a lack of innovative approaches to solve the complex and rapidly developing issues faced by governments today.

Of course, examples of digital democracy and online participation already exist (online petitions, suggestion systems, the use of social media sites to name a few) [5]. Yet, these tools tend to involve a relatively small and unrepresentative number of citizens and are often used for relatively marginal or predefined issues [2, 6]. The reasons for this are varied but include:

- An unwillingness of traditional government structures to adopt new, specific methods at scale – the sheer number of tools available dilute the public voice.
- A technological determinism – a linear or mechanistic assumption that technology is the solution to problems that actually require a socio-technical approach based on achieving an alignment between people's needs and technical affordances.
- A tendency for these online platforms to polarise opinions and circulate false information.

Many argue that government structures must not only embrace ICTs but, through the design of intelligent digital tools, include citizens more effectively within decision-making processes in ways that not only encourage greater participation and better decision making but also restore trust [3].

This paper describes ongoing research into one such tool, online platforms for public participation at the city-level. These are platforms set up by local governments to enable citizens to submit ideas and information, rank priorities and allocate public resources [2]. Two platforms, 'Decide Madrid' in Spain (https://decide.madrid.es/) and 'Better Reykjavik' in Iceland (https://betrireykjavik.is/) have so far been selected as case studies. Others will be added as the study progresses, selected to enable comparison of platform attributes and identification of their impact on observed outcomes.

The objective of this research is to determine if these platforms are successful in offering greater participation and better decision making or simply replicate the status quo. To be classed as successful, we argue that platforms must offer a new approach to those provided by existing models of digital democracy: platforms must take on a form of deliberative democracy, allowing for knowledge co-production and the continual emergence of a collective intelligence (which will be used as a measure of success). If these platforms are deemed to be unsuccessful, a new model of online participation will be suggested, adapted from traditional models of offline participation and the promise of collective intelligence.

The paper is structured as follows. Section 2 introduces the research questions and presents a definition of the platforms under investigation. Section 3 unpacks the theoretical background of citizen participation in democracy and the concept of collective intelligence. Section 4 introduces the methodology that will be used for this research project. The problems faced by the platforms under investigation are presented in section 5 as preliminary findings. Finally, after identifying the problems, section 6 suggests the development of a new model of online participation as a potential solution.

## 2. RESEARCH QUESTIONS

Before proposing our research questions, it is first necessary to define the city-level online platforms of participation. This is so we may differentiate them from the myriad of other tools available offering their own form of 'digital democracy'. This definition of city-level online platforms of participation, which to our knowledge has not yet been done, emerged from the preliminary literature review [2, 5] and has been used to select the case studies explored in greater detail in section 4. To be considered in this research the platform must:

- **Allow for an issue to emerge** – the online platform should take the form of a consultation forum where citizens are given the chance to present their ideas on issues regarding services and operations of the city.
- **Be created *for* the 'city level'** – the ideas and issues presented must directly relate to services and operations of the municipal district.
- **Be created *by* the city** – the platform itself must be created (or officially endorsed) by the relevant authority. As a result, the suggestions and comments made must, after a voting system, be formally addressed at the city council or government.
- **Be accessible to all** – all visitors to the site should be able to read the suggestions and comments. The submission of an idea, commenting or voting must either be open to all or made available through a simple registration confirming local residence (allowing only citizens of the city to directly participate).

Being relatively new, there has been no autonomous conclusion to whether or not these platforms are considered successful. Due to the absence of this evaluation, no framework has been provided on how to improve such platforms if they are deemed unsuccessful. Additionally, there has been no investigation into comparing and contrasting multiple online platforms from various cities to determine if a particular typology exists which explains their attributes. As a result, we propose the following inter-related research questions:

- **RQ1**: Do the online platforms offer or improve the formulation of questions and the negotiation and decision-making processes within heterogeneous and dispersed communities or simply offer, like many other tools available, a form of delusional democratisation of predefined participation?
- **RQ2**: How is success measured? If these platforms are successful, does a model exist that may be replicated elsewhere? If they are deemed unsuccessful how would one increase participation? What is the motivation for citizens to participate in the first place?
- **RQ3**: Does a form of collective intelligence emerge from these platforms (how people propose ideas, connect them, improve them, select the most relevant ones)? How does collective intelligence, when used for decision making, evolve as a practice?

### 2.1. Research impact

The conceptual pivot of this research is to determine if online platforms are successful in providing effective public participation in decision making. If so, do these platforms provide a model of digital democracy that may be replicated elsewhere? If unsuccessful, how might their deficiencies be addressed? The suggested outcomes will not only aid current platforms in improving the service they provide but may also be used as a guideline for future platforms launched by other organisations or city administrations.

## 3. LITERATURE REVIEW

As discussed above, many of the challenges faced by cities surpass the capabilities of traditional government structures. There is a need to harness the power of social and technological innovation to transform governance processes, models and practices. Though online platforms for public participation in decision making offer a range of new possibilities, they raise new concerns in respect to the level of participation they offer.

Before exploring and critiquing the current online platforms, it is first necessary to introduce the theoretical background of citizen participation and collective intelligence. we argue that although forms of participation exist across all levels of democracy, the concept of deliberative democracy (co-production as participation) offers the most effective form of citizen participation and therefore should be the model the online platforms strive to match.

### 3.1. Democracy and citizen participation

The debate on forms of democracy and appropriate levels of citizen participation in decision making is well established [8, 9].

According to Held [10] most forms of democratic theory can be divided into two broad types: representative democracy and direct democracy. We first explore these two forms, before delving deeper into the third model, deliberative democracy.

Representative democracy, where officials are elected to represent the best interests of citizens, has become established as the dominant form globally [3]. Yet many critics now believe this form of democracy to be outdated [3, 8, 10]. The emergence of ICTs allows for mass communication in technologically advanced societies, essentially dissolving the geographic distance that once limited a citizen's ability to represent themselves [9]. Additionally, the increasing complexity and speed of change make it almost impossible for decision makers alone to gather and understand all the information required to address contemporary challenges. Perhaps best summarised by Levy [3], the representative vote itself is a "molar process of social regulation". A citizen's identity is reduced to a binary obedience of 'yes' or 'no' to questions set by others on a four to five-year cycle. The response is ultimately used only for quantitative or statistical purposes, leaving no room for initiative. As he states, "the democratic ideal is not the election of representatives but the greatest *participation* of the people in public life." ([3] emphasis added).

The ideology of greater citizen participation is strongly related to the concept of 'direct democracy', which allows for a system of decision making in which citizens are directly involved [10]. Forms of direct democracy enable "everyone to help develop and refine shared problems on a continuous basis, introduce new questions, construct new arguments and formulate independent positions on a wide range of topics" [3]. Citizens' identify, rather than reduced to a statistic, would be defined by their contributions to the political landscape constructed by the mass, and continuously worked upon. In this way each identity would be unique but also coupled with the possibility of working with others having similar or complementary positions on a given subject or issue [3].

There is the danger here of prematurely concluding that, from its offering of collective action and the mobilisation of diverse skills, direct democracy is the most suitable model for combating the multifaceted problems faced by cities today. Indeed, widening participation to a form of direct democracy may provide democratisation and empowerment for the citizens involved yet there is a concern over structural barriers in society that may prevent such participation [11]. To critique the any case studies under this direct model would therefore be insufficient.

Online platforms, despite offering direct input, are often in danger of involving externally defined programmes of participation [12]. These programmes carry out pre-defined or fixed practices as a linear process, extended only to certain groups [13]. Concern is raised that this offers instrumentality (citizens as data providers) rather than empowerment (citizens opening up information spaces), giving what de Albuquerque and de Almeida [11] describe as a "delusion of democratization". Using a pedagogical lens, they proposed to embed participation through the process of knowledge co-production. Using Freire's *Pedagogy of the Oppressed* they reject the traditional 'banking model of education', where the teacher acts as the sole custodian of knowledge. This model prepares the individuals to give answers to pre-determined questions. Instead, they argue that the educator and learner should educate each other in a dialogical process while retaining asymmetrical character within. It should be noted that this asymmetry cannot be ignored as actors (government, citizens, scientists, policy makers) would never carry or play the same role. If actors' roles are assumed to be identical it becomes impossible to take into consideration the cultural background of the learner [11]. This thought is echoed by Levy [3] who states that a collective voice carries with it the danger of masking diversity and divergence. It may fail to integrate the differences that individualise the crowd.

A process of knowledge co-production will therefore fit a model of *deliberative democracy*. This type is broadly defined by Held [10] as a "learning process in and through which people come to terms with the range of issues they need to understand in order to hold a sound and reasonable political judgement… [and] a commitment to politics as an open-ended and continuous learning process in which the roles of the 'teacher' and 'curriculum' are raised and the matter of what is to be learnt is settled in the process of learning itself."

As Klein [14] argues, the *scale* of the deliberation process is essential. A small group of powerful stakeholders who create policies behind closed doors could be defined as a deliberative system. Yet, due to the scale and complexity of the problems faced today this is no longer adequate [14] and we argue would be closer to an epistocracy (a government run by only those with political knowledge) [5]. Instead, a move must be made from deliberative 'team' scales to 'crowd' scales. This will allow for access to a greater diversity of ideas, greater creativity and the production of high-quality results [14]. There is hope that crowd-scale, deliberative engagement will stimulate reflection, not only on the part of those immediately involved but also on the part of those who come into contact with deliberative activists – family members, friends, colleagues. This *re*-engagement of citizens in politics would stimulate widespread networking, which could trigger a culture of far-reaching civic participation [10].

In this sense a true form of citizen participation should be seen as a collective feedback loop or a form of dialogical communication at a crowd scale. Through the ongoing process of collective observing, listening, engaging, re-engaging, deliberating, evaluating, revising and adjusting through multiple stakeholders the online platform would utilise the crowds 'collective intelligence' in a form of knowledge co-production [3, 8].

### 3.2. Collective Intelligence

We have argued that participatory democracy gives potential for information and conclusions to form within a group, but it does not, so far, include feedback on a systematic and ongoing basis to permit the continual emergence of new insights that may affect other parts of the system. The 'collective intelligence' of a community, which is a complex, adaptive, self-organising and emerging system [15] would allow for such a concept and should therefore be considered as an essential component when investigating online platforms of participation.

Collective intelligence has been defined as a "form of universally distributed intelligence, constantly enhanced, coordinated in real time and resulting in the effective mobilisation of skills" [3] and "the general ability of a group to perform a wide variety of tasks" [16]. In its most basic sense, everyone knows something and there is no end point to this intelligence or knowledge transfer. Members of a community are able to coordinate their interactions by identifying their diverse skills allowing for the co-production of ideas [3, 16, 17].

In a detailed review on what collective intelligence means in a human context, Salminen [17] finds three levels of abstraction: the micro-level, the macro-level and the level of emergence that resides between the two. At a micro-level, behavioural features such as trust, intelligence and motivation are the enabling factors of collective intelligence. The immersion within a social network is a typical human condition and our ability to understand and exhibit social signs allows for smooth coordination with this network [17]. Pentland [18] argues that key elements of human intelligence could reside in network properties, which Woolley et al. [16] supports. Using a wide variety of cognitive tasks traditionally used to measure individual intelligence (factor $g$), Woolley et al. [16] systematically examined whether a general collective intelligence factor exists in groups. In two studies they found evidence of a '$c$ factor' that explains a group's performance on such tasks and can be used to predict performance on other tasks. Interestingly this factor is not strongly correlated with the average or maximum individual intelligence of group members ($g$) but is correlated with; the 'average social sensitivity' of group members if conversational turn taking is equally distributed, and the proportion of females in the group (which the authors conclude is mediated by social sensitivity, as women scored better on this measure).

At a macro-level, collective intelligence becomes a statistical phenomenon through the 'wisdom of crowds' based on diversity, independence and aggregation [17]. Diversity refers to differences in demographic, educational and cultural backgrounds and also the way people solve problems [17]. Both simulations [19] and experiments [20] have shown that under certain conditions groups of diverse problem solvers can outperform groups of high-ability problem solvers. Independence is when individuals are not influenced by others in the network, which may reduce the group's diversity (similar to retaining asymmetry within the dialogical process). Aggregation combines and processes individual estimations into a collective one [17].

System behaviour at the macro-level emerges from interactions between individuals at the micro-level. Theories of complex adaptive systems are used to explain how collective intelligence as a statistical phenomenon emerges from individual interactions [15, 17]. Complex systems are characterised by adaptivity (changing according to the milieu), self-organisation (order without central control) and emergence (whole is more than the sum of its parts) [15, 17].

Based on these factors, micro-level features (intelligence, trust, motivation) are the enabling factors of collective intelligence and set humans apart from other collective intelligence systems (for example, motivation is not considered in algorithms). Individuals interacting with one another form a complex adaptive system that shows self-organisation and emergence. At the macro level this system becomes probabilistic with diversity, independence and ways to aggregate information becoming important features of the system [17]. From this we may:

- Determine if collective intelligence emerges from the online platforms by:
    - Measuring the macro-level features to predict the performance of the system as a whole.
    - Understand how the micro-level activities lead to macro-level behaviour.
- Develop a common framework or model capable of explaining how collective intelligence works and therefore how it may be measured and engineered (rather than simply noting its emergence) [15].
- Determine if a group's collective intelligence could be improved by better digital collaboration tools.

## 4. METHODOLOGY

Figure 1 provides a summary of the methodology for the study. First, a preliminary literature review was carried out to establish the research questions and shape a definition for city-level online platforms of participation. Relevant case studies were then identified from this definition (phases one and two, Figure 1). From an initial observation of these platforms a number of existing challenges were recognised (see section 5.2), suggesting that current platforms are not successful in offering greater participation and better decision making.

The next stage of research (phases three and four, Figure 1) will explore the causes of these problems and build upon the initial observations. The final stage of research (phase five, Figure 1) will collate and analyse all findings and offer a potential solution in the form of a new model of online participation.

### 4.1. Tracking of a successful proposal

There is a danger to deem a proposal successful once it has received enough support votes. Yet, the proposal itself must be enacted upon within the relevant government. Through an ethnographic lens, we will follow a successful proposal from its generation to outcome in order to develop a deeper understanding of platform processes, goals, the actors involved and how this may differ from the traditional, offline route.

### 4.2. Manipulation of online platform data

The Alan Turing Institute, in collaboration with Warwick University, recently won a bid through the innovation foundation Nesta to investigate the effectiveness of the Decide Madrid Platform. Through this collaboration we have been granted access to the proposals and comments made on the platform (over 430,000 users submitting 23,000 proposals, 170,000 comments and more than 3,000,000 votes of support). We will use this data to determine the type of proposals submitted to the platform. From this we may infer what the users of the platform deem to be the more important issues faced by the city.

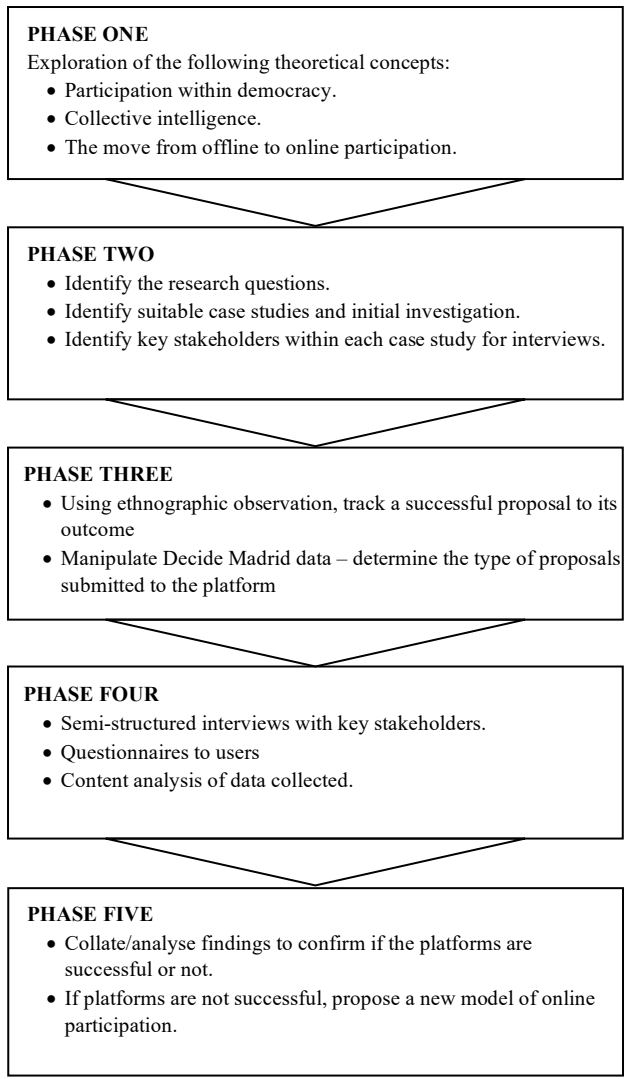

**Figure 1: Overview of methodology**

## 4.3. Interviews with case study stakeholders

We will organise interviews with stakeholders for each case study (including platform developers, community members, city officials and administrators). The main purpose of the interview would be to explore meaning and perceptions to gain a better understanding of each project (its origins, how the platform works, the actors involved).

Interviews will follow a semi-structured format where knowledge is not given but encountered through conversation [21, 22]. A semi-structured interview is organised around a set of predetermined open-ended questions, with other questions emerging throughout [23]. This type of interview not only allows for following up on important issues but also allows for a more dialogical process of knowledge co-production. As Bernard [24] states, this format provides an efficient use of the interviewees' time and therefore works well with "high-level bureaucrats and elite members of a community" (p158), two groups we may well encounter in this research.

Preferably the interviews will be conducted face-to-face to allows for interpersonal contact and conversational flexibility [22, 24]. Where case study location makes this impractical, interviews will be conducted by telephone [25] or online [26].

The analysis of data will occur alongside this data collection, so that we may generate an emerging understanding about the research question, which in turn informs both the sampling and the questions being asked [23]. Following this, we will apply Hermeneutic Content Analysis (HCA) for interpretation [27]. HCA moves beyond the traditional quantitative and qualitative content analysis of systematic coding and describing [24, 28] to also include understanding and reflection through methods of interpretation and reinterpretation [29]. By *understanding*, this method avoids objectively observing and analysing non-numerical data and allows for me to relate this data to cultural, historical, political and social contexts.

## 4.4. Questionnaires to users of the online platform

In addition to interviews with key stakeholders, we will design and distribute a questionnaire to users of the online platform. This will be to determine how the platform under investigation is used and whether there are significant differences in resources, attitudes, and skills of the users [6]. To expand upon this, we will also determine if micro-level features (intelligence, trust, motivation) that enable collective intelligence to emerge [17] exist within this community. We will also measure macro-level features (particularly diversity and independence) to determine the performance of the platform as a whole.

## 5. PRELIMINARY FINDINGS

### 5.1. Case studies

Two case studies have been selected so far following the definition of city-level online platforms in section 2. The platforms have similar attributes, allowing for a comparison of features and issues faced and a consistent methodological approach. More case studies will be identified as the study progresses allowing for a stronger comparison.

#### 5.1.1. *Decide Madrid (Madrid, Spain)*

Decide Madrid was launched as a result of the 2011 15-M (anti-austerity) demonstrations in Spain. This movement was aided by a dense online network of activists and bloggers campaigning and sharing via social media. A minority government was formed soon after the protests and in September 2015 the city of Madrid launched 'Decide Madrid' with the intention of "promoting more direct democracy, accountability and transparency in local decision-making" [2].

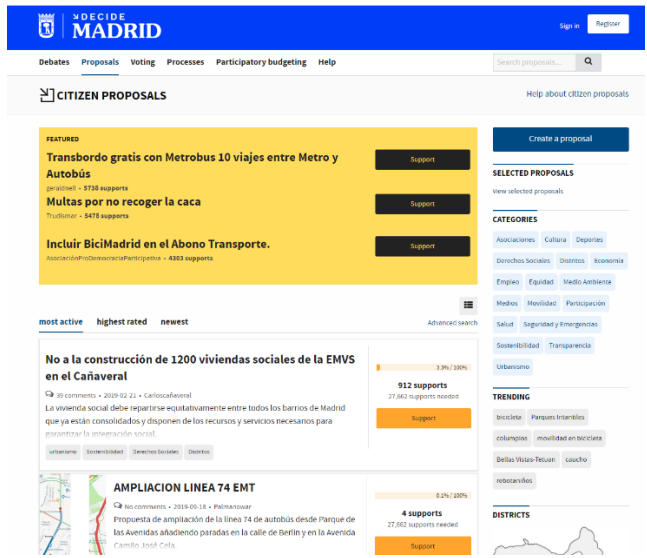

Figure 2: Screenshot of the Decide Madrid platform

The online platform (Figure 2) can be divided into four components: proposals and votes, debates, participatory budgeting and consultations. Within the proposals and votes section, a resident may create a proposal for a new local law which is shared on the platform for 12 months. During this time other residents may comment on the proposal or cast votes of support. If the proposal gains approval from 1% of the census population with the right to vote (the equivalent of 27622 supporters) it is positioned at the top of the webpage. The proposal is then debated for another 45 days before going to a final public vote. If this is approved, the Council has one month to draw up technical reports on the legality, feasibility and cost of the proposal which are all published on the platform [33].

*5.1.2. Better Reykjavik (Reykjavik, Iceland)*
The 2008 Icelandic financial crisis and subsequent political crisis resulted in a dramatic fall in trust in Parliament (from 40% in 2008 to 11% in 2011) and eventual replacement of the Government [31]. The newly formed 'Best Party' won local council elections in Reykjavik and promised greater transparency, new political actors and a greater role for citizens in decision making. In the same year a non-profit organisation, the Citizens Foundation, developed 'The Shadow City', an open-source crowdsourcing online tool, which enabled users to discuss innovative ideas at the city level. As the 'Best Party' did not have a conventional manifesto, they embraced 'The Shadow City' as a tool for generating citizen-led ideas. In 2010 they asked the platform founders to set up a specific platform for the council - Better Reykjavik (Figure 3) [2].

Registered users participate by suggesting, debating and rating ideas for improving Reykjavik. Responses are separated in columns, either for or against the idea (which makes it difficult to reply directly to someone you disagree with and also helps to show the multiplicity of views on a subject). Both ideas and comments can be voted on by the online community. At 12:00pm on the last weekday of every month, the 5 most popular ideas are reviewed by Reykjavik City Council. The council's response is published on the platform.

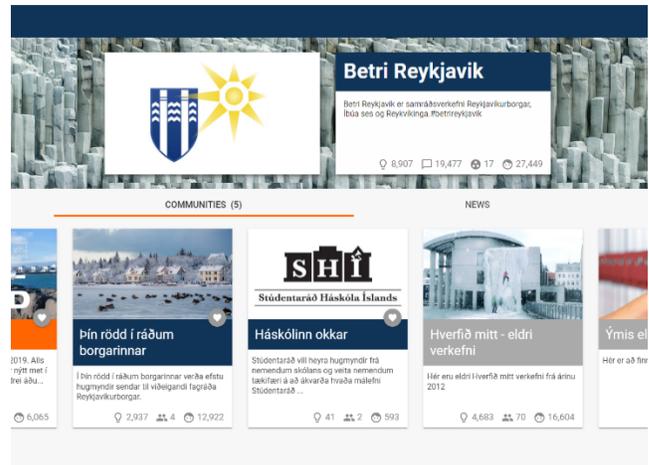

Figure 3: Screenshot of the Better Reykjavik platform

## 5.2. Issues faced by current online platforms

Following a brief review of each case study, the following issues were identified. These are only preliminary findings, but they are nevertheless valuable for establishing the foundations for the later stages of the study.

*5.2.1. Participation*
For each of these projects, raising awareness and engagement remains a challenge. For example, in Decide Madrid, 56% of Madrid's residents are aware of the platform's existence and less than 10% of the city's population is registered. Despite halving the number of votes required for approval (from 2% of the census population to 1%) only 2 out of over 13,000 proposals have ever made it through [2].

The reasons for lack of participation vary. They include lack of awareness about the platforms, a lack of online access, a lack of time or interest, concerns about the quality of proposals, fear the budget available is too small to create any meaningful change and that more pressing socio-economic projects were not being addressed. Many citizens, although having an individual interest, could abstain from collective action by simply free riding on others' contributions (a participation dilemma) [32]. It could also be argued that the sheer number of platforms of deliberation and digital democracy (online petitions, suggestion systems, e-discussion forums, etc.) has diluted the public voice [5].

In addition to poor participation, *diverse* participation also remains an issue. For example, a recent evaluation of the Better Reykjavik platform found that participation is biased towards those that are University educated, those who have higher salaries and people aged between 36 and 55 [2]. More work must be done to increase participation from under-represented groups to highlight more pressing issues faced within cities.

*5.2.2. Generating ideas to address more complex issues*

Many suggested ideas relate to aesthetic or minor environmental improvements, such as tree plantation or graffiti removal rather than initiatives to tackle more complex issues. Although all ideas suggested should be considered, more work is needed on how citizens could be directed to tackle more complex issues, how to utilise the collective intelligence of the crowd efficiently and how government officials could engage with residents who have the highest barriers to civic participation (a blend of top-down and bottom-up approaches). Barriers to addressing more complex issues include a lack of public understanding of the platform's process and the scope of city powers, a fear (again) that the budget available is too small to create any meaningful change [2], and, importantly, the structure of the platform itself.

Addressing more serious and complex city issues requires a continuous network approach that may be achieved under a form of deliberative democracy. The current forms of online participation still use the traditional democratic model of a fixed or finalised proposal (ideas are made over a time frame and submitted to the local council as a finalised idea), which encourages superficial suggestions. This gives the impression that the issue itself is finite, with a start and end and logical structure to which we may apply previous systems of control to and evaluate. For more complex issues there is never a final point where the problem is 'solved', rather it constantly needs acting upon. A more fluid model and platform structure is therefore needed.

*5.2.3. Visibility*

Despite poor participation rates, these online platforms still receive a large number of proposals from their active members [2]. Yet, this high volume often makes it difficult for users to identify proposals of interest, leading to a high degree of duplication. With so many ideas, comments and suggestions being made, it is increasingly challenging to find the optimal balance between providing citizens with the level of information and details they need to make informed decisions (without overwhelming them) [5]. This issue relates to the lack of participation (due to concerns about the quality of proposals made) and the structure of the platform itself.

*5.2.4. Resources*

Despite claiming to be a form of collective participation, the large number of ideas generated by citizens must still be processed by civil servants, placing a strain on both city officials and IT systems. It takes time, money and proactive outreach to run such a programme at city-scale. For example, €200,000 of public money was used to promote the 2016 Decide Madrid process [33], which led to a spike in activity that was not sustained.

## 6. CONCLUSIONS AND FURTHER WORK

Underlying theoretical concepts, case studies and challenges have now been identified. Our initial investigations of two case studies reveals a number of issues prevent deliberative democracy taking root. Drawing on these findings, we have developed a preliminary framework of factors for the emergence and sustainability of collective intelligence at the macro-level (Figure 4).

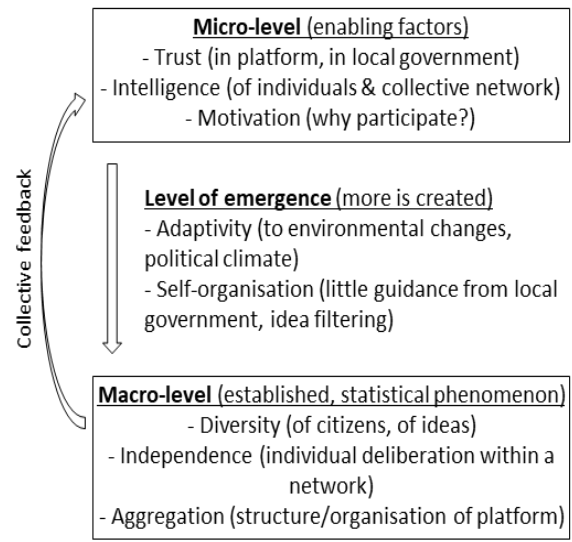

**Figure 4: Collective intelligence framework**

The final stage will collate and analyse the data (ethnographic observation, platform activity, questionnaires and interviews) in order to test and extend these initial findings. Then, we will use the results and the framework to develop a model for sustainable online participation, adapted from traditional models of offline participation [6].

The model will also consider new online-related resources and citizen experiences, ideological and collective motives and selective incentives. It is hoped that this will encourage a form of mobilisation (reaching out to new, formally inactive groups) and empowerment of citizens rather than a normalisation of online political activity through citizens-as-instruments.

## ACKNOWLEDGMENTS

This research is funded by the Engineering and Physical Sciences Research Council (EPSRC) (grant no. EP/L016400/1) for the EPSRC Centre for Doctoral Training in Urban Science, University of Warwick.